\documentclass[twocolumn,showpacs,preprintnumbers,amsmath,amssymb]{revtex4}

\usepackage{amsmath}
\usepackage{graphicx}
\usepackage{color}
\usepackage{dcolumn}% Align table columns on decimal point
\usepackage{bm}% bold math

\def\beq{\begin{eqnarray}}
\def\eeq{\end{eqnarray}}
\def\bea{\begin{eqnarray}}
\def\eea{\end{eqnarray}}

\begin{document}

\preprint{DCP-07-01}

\title{{Neutrino mixing from the double tetrahedral group $T^{\prime}$}}
\author{Alfredo Aranda}
 \email{fefo@ucol.mx}

\affiliation{Dual C-P Institute of High Energy Physics}
\affiliation{Facultad de Ciencias, Universidad de Colima,\\
Bernal D\'{i}az del Castillo 340, Colima, Colima, M\'exico}

\date{\today}

\begin{abstract}
It is shown that it is possible to create successful models of
flavor for both quarks and leptons using the discrete non-abelian
group $T^{\prime}$ by itself. Two simple realizations are presented
that can be used as the starting point for more general scenarios.
In addition to the Minimal Supersymmetric Standard Model particle
content, the models include three generations of right handed
neutrinos and four scalar flavon fields. Three of the flavons are
needed in the quark and charged lepton sector of the models and the
fourth flavon participates only in the neutrino sector.
\end{abstract}


\maketitle

The recent results obtained by experiments in neutrino
physics~\cite{neutrinoresults} have reawakened the need to
understand the rich structure of masses and mixing angles patterns
of fundamental particles. More so when the results regarding
neutrino mixing turned out to be in the range less expected by most
model builders (biased perhaps by the mixing in the quark sector).

The addition of flavor symmetries to the structure of the Standard
Model (SM) constitutes one of the main venues that have been used to
explore the situation~\cite{Ross:2000fn}. In particular the use of
discrete symmetries has been known to facilitate the creation of
elegant and simple models of flavor that can easily accommodate the
patterns observed in the quark and charged lepton
sectors~\cite{mareview,altarelli,deMedeirosVarzielas:2006fc}, and in
view of the recent results in the neutrino sector, several attempts
have been made to incorporate them. In fact, if one is interested in
the lepton sector alone, the use of the $A_4$ group as a flavor
symmetry group stands out. It is possible to generate the so-called
tribimaximal mixing matrix in a natural way and several specific
models in the literature make use of this fact~\cite{A4papers}. The
crucial aspect of this success is the existence of a three
dimensional representation in $A_4$ which can be used to {\it group}
the three neutrinos. The models however cannot incorporate the quark
sector in a natural way and if one is interested in accomplishing
this then an extension of the group is necessary.

One group that shares the properties of $A_4$ is the bigger double
tetrahedral group $T^{\prime}$\cite{frampton1}. The key virtue of this group is that
in addition to the three-dimensional representation, $T^{\prime}$
contains two-dimensional representations that can be used in the
quark sector. Flavor models that use $T^{\prime}$ (in direct product
with other abelian factors) have been used before and were able to
successfully account for the patterns observed in the quark sector,
however these models presented a solution for the neutrino sector
that corresponded to the now excluded small mixing angle
solution~\cite{maximal}. Motivated by these issues there have been
some recent proposals that use $T^{\prime}$ in order to extend the
$A_4$ models~\cite{feruglio,muchunchen}. In \cite{feruglio} a
supersymmetric model has been presented with the flavor symmetry
$T^{\prime}\otimes Z_3\otimes U(1)_{FN}$ where the $T^{\prime}$ part
of the model is responsible for the maximal mixing in the lepton
sector. In~\cite{muchunchen} a model was presented that uses
$T^{\prime}$ together with a $Z_{12}\times Z^{\prime}_{12}$ as
flavor symmetries in the context of SU(5) unification. Other works
using $T^{\prime}$ in similar contexts can be found
in~\cite{frampton2}.

In this letter we show that it is possible to create a simple flavor
model that incorporates both quarks and leptons using only
$T^{\prime}$ as a flavor symmetry. Furthermore we show that this is
possible with a minimal set of additions in the flavor sector. The
purpose of this analysis it to provide a simple framework that can
be used as a starting point for more general extensions. In order to
set it we summarize the assumptions we make as follows:
\begin{itemize}
    \item We work in the context of the Minimal Supersymmetric Standard Model (MSSM).
    \item Three heavy right handed neutrinos are present in the model.
    \item We assume that the light neutrino masses are generated via
    the seesaw mechanism.
    \item All Yukawa couplings are naturally of O(1).
    \item We assume the flavor symmetry to be a global symmetry.
\end{itemize}

Perhaps with the exception of the last two, these assumptions are
considered to be naturally expected in most extensions of the
Standard Model. Regarding the requirement of O(1) coefficients, this
amounts to the fact that we are trying to understand the huge
differences in scales of the different masses of the fundamental
particles, and so, we adhere to expectation that all Yukawa
couplings must naturally be of O(1). The observed differences in
masses and mixing angles must then be produced by the flavor
structure.

We are after a simple framework that can easily be incorporated in
more general scenarios of physics beyond the Standard Model. Thus,
even when we have strong reasons to suspect that global symmetries
are inconsistent as fundamental symmetries of nature, they suffice
at the level we are working.

Given the success of $T^{\prime}$ as a discrete symmetry of flavor
in the quark sector, we choose to explore the possibility of
creating a complete model based on this symmetry. $T^{\prime}$ is
defined as the group of all 24 proper rotations in three dimensions
leaving a regular tetrahedron invariant in the SU(2) double covering
of SO(3). It has three singlets ${\bf 1^0}$ and ${\bf 1^{\pm}}$,
three doublets, ${\bf 2^0}$ and ${\bf 2^{\pm}}$, and one triplet,
${\bf 3}$. The triality superscript provides a concise way of
stating the multiplication rules for these reps: With the
identification of $\pm$ as $\pm 1$, trialities add under addition
modulo three, and the following rules hold: \beq \label{rules}
\nonumber {\bf 1} \otimes {\bf R} &=& {\bf R}\otimes {\bf 1} \ \
{\rm for \ any \ rep}
 \ {\bf R}, \\ \nonumber  {\bf 2}\otimes {\bf 2} &=& {\bf 3} \oplus {\bf 1} \\
\nonumber {\bf 2} \otimes {\bf 3} &=& {\bf 3} \otimes {\bf 2}={\bf
2^0}\oplus {\bf 2^+}\oplus {\bf 2^-}, \\ {\bf 3} \otimes {\bf
3}&=&{\bf 3}\oplus {\bf 3}\oplus {\bf 1^0}\oplus {\bf 1^+}\oplus
{\bf 1^-} \ . \eeq

A nice way to describe this group is the following: the group of all
proper three dimensional rotations that leave a tetrahedron
invariant is called the tetrahedral group and it is denoted by $T$.
It is easy to show that it has $12$ elements. One can construct it
by parameterizing the group SO(3) of all proper three dimensional
rotations in terms of Euler angles, and then restricting to the
specific values describing rotations that leave the tetrahedron
invariant. Those Euler angles also describe rotations in SU(2) space
and so $T^{\prime}$ is the subgroup of SU(2) corresponding to the
same Euler angle as $T$ in SO(3). This has as a consequence that
even-dimensional representations of $T^{\prime}$ are spinorial while
odd ones coincide with those of $T$. For complete details regarding
the group structure of $T^{\prime}$ see~\cite{maximal}.

As mentioned above we use $T^{\prime}$ as a global symmetry of
flavor~\footnote{As described in~\cite{maximal} $T^{\prime}$ can in
principle also be used as a local symmetry of flavor however not by
itself.}. In~\cite{maximal} it was shown that it is possible to
reproduce the observed patterns of masses and mixing angles in the
quark and charged lepton sectors using $T^{\prime}$. This is
accomplished first by choosing the following assignments:
\begin{equation}
\psi \sim {\bf 2}^- \oplus {\bf 1}^0 \,\,\, \mbox{ for } \,\,\,
\psi=Q, U, D, L \mbox{ and } E,
\end{equation}
for matter fields and $H_{U,D} \sim {\bf 1}^0$ for the MSSM Higgs
fields. This yields
\begin{eqnarray} \label{Yuk}
Y_{U,D,L} & \sim & \left( \begin{array}{c|c} [ {\bf 3}\oplus {\bf
1}^{-}]
& [ {\bf 2}^{+}] \\
\hline {[} {\bf 2}^{+}]
 & [ {\bf 1}^{0} ]
\end{array} \right) \ .
\label{eq:ytp}
\end{eqnarray}
We then introduce the three flavons $\phi$, $S$ and $A$ transforming
as ${\bf 2}^{+}$, ${\bf 3}$ and ${\bf 1}^{-}$, respectively, and
with vacuum expectation values (vevs) given by
\begin{equation}
\frac{\langle\phi\rangle}{M_f} = \left(\begin{array}{c} 0
\\ \epsilon\end{array}\right),
\ \ \frac{\langle S\rangle}{M_f} = \left(\begin{array}{cc} 0 & 0
\\ 0 & \epsilon \end{array}
\right), \ \  \frac{\langle A\rangle}{M_f}= \left(\begin{array}{cc}
0 & \epsilon'
\\ -\epsilon' & 0 \end{array} \right),
\end{equation}
where $M_f$ is denotes the flavor scale and where
$\epsilon = 0.02$ and $\epsilon^{\prime}=0.002$.
Note that there is a two-step sequential breaking of the symmetry where
\begin{eqnarray} \label{eq:nbreak}
T'  \stackrel{\epsilon}{\longrightarrow} Z_{3} \stackrel{\epsilon'}
{\longrightarrow} nothing.
\end{eqnarray}
The vev choices above are an assumption at this level and we consistently
assume that any flavon with nontrivial transformation properties under
$T^{\prime}$ and the residual $Z_3$ either gets a vev of the same order
the breaking of either symmetry or gets no vev at all.

From Eq.(\ref{Yuk}) it can be seen that there is no explanation at this
level of the ratio $m_t/m_b$. Grand unified versions of this model can be
constructed that generate such ratio in a natural way. We use the simplest
version given above where an overall free parameter $\xi\approx 0.01$
multiplies both $Y_D$ and $Y_L$ in order to account for it~\cite{maximal}.
Introducing O(1) parameters in all of the entries of the matrices in 
Eq.(\ref{Yuk}), it is possible to reproduce all quark masses and CKM mixing
angles~\cite{barbieri,maximal}.

If we insist in using only two and one-dimensional representations
for the matter fields, we have three possibilities for the
neutrinos. The first one is to simply keep the same format and
assign them in the ${\bf 2} \oplus {\bf 1} $ fashion. Unfortunately
this does not lead to good phenomenology~\cite{maximal}. There are
two more possibilities: We can use the assignment ${\bf 1} \oplus
{\bf 2}$, i.e. we {\it group} into a doublet the right-handed
neutrinos of the second and third generations (we call this set A).
The last choice is to {\it group} into a doublet the first and third
generation of right-handed neutrinos: $({\bf 2})_{1} \oplus {\bf 1}
\oplus ({\bf 2})_2$ (set B). An interesting observation is that
since the right-handed neutrinos are only charged under the
additional flavor symmetry, one naturally expects that the three
possibilities mentioned above are equivalent up to a redefinition of
their generation. However, the flavor structure does communicate
this generation information and hence differentiate between the
different sets through the seesaw mechanism and through the flavon
fields. This is implicitly manifested in the fact that the ${\bf 2}
\oplus {\bf 1} $ assignment for right-handed neutrinos does not
work~\footnote{We thank A. Smirnov for pointing this out.}. We now
discuss each possibility separately:

{\bf Set A}: We assign the three right-handed neutrinos to the
representation ${\bf 1}^-\oplus 2^-$ and introduce a new flavon
field $\phi_{\nu}\sim {\bf 2}^{-}$ with vev given by $\langle
\phi_{\nu} \rangle ^T = (\epsilon \ \ \epsilon^{\prime})$. Note that
this is the only addition we make and that this flavon does not
couple to the quark nor to the charged lepton sector. Given these
assignments we obtain the following textures (at leading order):
\beq M_{LR} & \sim & \left(
\begin{array}{c|c} [ {\bf 2}^-]
& [ {\bf 3}\oplus {\bf 1}^-] \\
\hline {[} {\bf 1}^{+}]
 & [ {\bf 2}^{+} ]
\end{array} \right) \rightarrow
\left( \begin{array}{ccc} l_1 \epsilon
& 0 & l_2 \epsilon^{\prime} \\
l_1 \epsilon^{\prime} & - l_2 \epsilon^{\prime} & l_3 \epsilon \\ 0
& 0 & l_4 \epsilon \end{array} \right) \langle H_U \rangle \\
M_{RR} & \sim & \left( \begin{array}{c|c} [ {\bf 1}^- ]
& [ {\bf 2}^{-}] \\
\hline {[} {\bf 2}^{-}]
 & [ {\bf 3}^{-} ]
\end{array} \right) \rightarrow
\left( \begin{array}{ccc}  \ r_1 \epsilon^{\prime}
& r_2 \epsilon & r_2 \epsilon^{\prime} \\
r_2 \epsilon & 0 & 0 \\ r_2 \epsilon^{\prime} & 0 & r_3 \epsilon
\end{array} \right) \Lambda_R \ ,
\eeq where we explicitly show all the O(1) coefficients that are
present in each matrix element. These O(1) coefficients are
important when a detailed numerical analysis is performed and one
needs to guarantee the results to be independent of these
parameters, i.e. one needs to verify that the results obtained are
due to the structure generated by the flavor symmetry and not by
coincidental cancellations and/or enhancements of combinations of
O(1) coefficients.

Using the seesaw formula $M_{LL}\approx M_{LR}M_{RR}^{-1}M_{LR}^T$
we obtain
\begin{eqnarray}
M_{LL}  \sim  \left( \begin{array}{ccc} (\epsilon'/\epsilon)^2 &
(\epsilon'/\epsilon)^3 & \epsilon'/\epsilon
\\ (\epsilon'/\epsilon)^3 & 1 & 1
\\ \epsilon'/\epsilon & 1 & 1
\end{array} \right) \frac{\langle H_U \rangle^2 \epsilon}{\Lambda_R} \ ,
\end{eqnarray}
where we have omitted the O(1) coefficients for clarity. Note that
the texture of $M_{LL}$ has the general form of the mass matrix that
leads to maximal mixing angle for the atmospheric neutrinos and LMA
solution for the solar neutrinos~\cite{muchun2} as desired.

In order to verify that this texture does reproduce the observed
patterns in the lepton sector, we diagonalize both $M_{LL}$ and
the charged lepton mass matrix in order to obtain the lepton masses and the CKM-like
mixing matrix of the lepton sector $ V = U_{L}^{\dagger} W$, where
$M_{LL}^D=W^{\dagger} M_{LL} W$ and $Y_{L}^D=U_L^{\dagger} Y_L U_R$.

From Eq.(\ref{Yuk}) we explicitly obtain the expression for the charged leptons
mass matrix (including O(1) coefficients)
\beq \label{YL} Y_{L} & \sim & \left(
\begin{array}{c|c} [ {\bf 3}\oplus {\bf 1}^{-}]
& [ {\bf 2}^{+}] \\
\hline {[} {\bf 2}^{+}]
 & [ {\bf 1}^{0} ]
\end{array} \right)
\approx\left( \begin{array}{ccc}
0 & l_5 \epsilon' & 0 \\
- l_5 \epsilon' & l_6 \epsilon & l_7 \epsilon \\
0 & l_7 \epsilon & l_8
\end{array} \right)\xi.
\eeq

Using these mass matrices one can easily find sets of O(1) coefficients that lead 
to the correct experimental results and
a full systematic analysis incorporating both a $\chi^2$
minimization analysis and the running of gauge and Yukawa couplings will be presented
elsewhere. 

We emphasize that the same 
procedure is performed in the quark sector and that the results for the quark masses and CKM mixing
angles are in perfect agreement with data as can be seen in Tables 3 and 5 of~\cite{barbieri} 
and Tables III and IV of~\cite{maximal}. Our quark sector is the same because the new flavon 
introduced in the neutrino sector does not alter the textures in the quark sector.

{\bf Set B}: We now assign the three right-handed neutrinos to the
representation $({\bf 2})^-{1}\oplus 1^- \oplus ({\bf 2})^-_2$ and
introduce the same flavon field $\phi_{\nu}\sim {\bf 2}^{-}$ with
vev given by $\langle \phi_{\nu} \rangle ^T = (\epsilon \ \
\epsilon^{\prime})$. We now obtain the textures: \beq M_{LR}
\nonumber & \sim & \left(
\begin{array}{c|c|c} [ {\bf 3}_{11}\oplus{\bf 1}^-_{11}]
& [ {\bf 2}^-_1] & [{\bf 3}_{12}\oplus {\bf 1}^-_{12}] \\
\hline [ {\bf 3}_{21}\oplus {\bf 1}^-_{21}]
 & [ {\bf 2}^{-}_2]& [{\bf 3}_{22}\oplus {\bf 1}^-_{22}] \\
 \hline [{\bf 2}^+_1] & [{\bf 1}^+] & [{\bf 2}^+_2]
\end{array} \right) \\
&\rightarrow & \left( \begin{array}{ccc}
0 & l_1 \epsilon & l_2 \epsilon^{\prime} \\
-l_2\epsilon^{\prime} &  l_1\epsilon^{\prime} & l_3\epsilon \\
0 & 0 & l_4\epsilon \end{array} \right) \langle H_U \rangle \\
\nonumber
M_{RR} & \sim & \left(
\begin{array}{c|c|c} [ {\bf 3}_{11}]
& [ {\bf 2}^-_1] & [{\bf 3}_{12}] \\
\hline [ {\bf 2}^{-}_{1}]
 & [ {\bf 1}^{-}]& [{\bf 2}^-_{2}] \\
 \hline [{\bf 3}_{21}] & [{\bf 2}^-_2] & [{\bf 3}_{22}]
\end{array} \right) \\
& \rightarrow & \left(
\begin{array}{ccc}
0 & r_1\epsilon & 0 \\
r_1\epsilon & r_2\epsilon^{\prime} & r_3\epsilon^{\prime} \\
0 & r_3\epsilon^{\prime} & r_4\epsilon
\end{array} \right) \Lambda_R \ .
\eeq
Note that in this case $M_{RR}$ has one extra O(1) coefficient compared
with the same matrix in the case A. This is due to the way the
flavons couple to the different entries and makes explicit our comment
above regarding the differences in both scenarios.

Using the seesaw formula we obtain (omitting O(1) coefficients)
\begin{eqnarray}
M_{LL}  \sim  \left( \begin{array}{ccc} (\epsilon'/\epsilon)^2 &
(\epsilon'/\epsilon)^3 & \epsilon'/\epsilon
\\ (\epsilon'/\epsilon)^3 & 1 & 1
\\ \epsilon'/\epsilon & 1 & 1
\end{array} \right) \frac{\langle H_U \rangle^2 \epsilon}{\Lambda_R} \ ,
\end{eqnarray}
i.e. the same texture obtained before. 

It is remarkable that with the addition of a single flavon in the
neutrino sector we obtain the right textures for $M_{LL}$ in both
cases. Assigning specific values to all the O(1) coefficients in
$M_{LL}$ one can easily (there are many possible sets) reproduce
the observed mixing angles and the
ratio of mass squared differences. As mentioned before, we have
presented two particular examples as an illustration of the success of the
models and will present the complete
numerical analysis including the running of gauge and
Yukawa couplings from the flavor scale down to the
electroweak scale as well as a detailed
$\chi^2$ minimization procedure elsewhere.

In this letter we have only considered the case where the matter
fields are assigned to singlets plus doublets of $T^{\prime}$. We
have done this based on a {\it argument} of naturalness, i.e quarks
and charged leptons clearly follow such structure and one might
expect neutrinos to do the same. On the other hand one can argue
that neutrinos tell us that there are big differences and therefore
one must keep an open mind. One can certainly choose to assign the
neutrinos to the three different singlet representations ${\bf 1}^0$
and ${\bf 1}^{\pm}$. We however are not interested in this
possibility because it is less restrictive, i.e. the flavon
structure from such assignments would require introduction of more
flavons and independent parameters, and so even when technically
feasible, we do not find it very revealing. Another option is to
assign them to the ${\bf 3}$. This is certainly possible and one can
in fact explore different possibilities, for example by assigning
both the right-handed neutrinos and the lepton SM fields to ${\bf
3}$'s, or leaving out the right-handed neutrinos altogether (and
thus forgetting about our assumption regarding the seesaw). These
possibilities are currently under study.

A final interesting observation is that even though SU(2) cannot be
used as a flavor symmetry in supersymmetric models (unless scalar
universality is assumed), its discrete subgroup $T^{\prime}$ can. It
is due to the fact that there are several doublet representations as
opposed to the single one of SU(2).

In conclusion we have shown that it is possible to create successful
models of flavor using $T^{\prime}$ and only $T^{\prime}$ as a
global flavor symmetry for both quarks and leptons. Working in the
context of the MSSM with three right-handed neutrinos and assuming
that the light neutrino masses are obtained through the seesaw
mechanism, we have presented two simple realizations that can be
used in more general contexts. In order to obtain the right
relations for masses and mixing angles in the quark and lepton
sectors, four new scalar flavon fields were introduced, one of which
only participates in the neutrino sector of the models.

\begin{acknowledgments}
We thank the Aspen Center for Physics for its hospitality while this
work was in progress. We thank Ernest Ma, Paolo Amore and Lorenzo
Diaz for very useful discussions. This work was supported in part by
PROMEP and by CONACYT.
\end{acknowledgments}


\begin{thebibliography}{99}
\bibitem{neutrinoresults}
   Q.~R.~Ahmad {\it et al.}  [SNO Collaboration],
  %``Direct evidence for neutrino flavor transformation from neutral-current
  %interactions in the Sudbury Neutrino Observatory,''
  Phys.\ Rev.\ Lett.\  {\bf 89}, 011301 (2002)
  [arXiv:nucl-ex/0204008];
  %%CITATION = PRLTA,89,011301;%%
  C.~K.~Jung, C.~McGrew, T.~Kajita and T.~Mann,
  %``Oscillations of atmospheric neutrinos,''
  Ann.\ Rev.\ Nucl.\ Part.\ Sci.\  {\bf 51}, 451 (2001);
  %%CITATION = ARNUA,51,451;%%
  K.~Eguchi {\it et al.}  [KamLAND Collaboration],
  %``First results from KamLAND: Evidence for reactor anti-neutrino
  %disappearance,''
  Phys.\ Rev.\ Lett.\  {\bf 90}, 021802 (2003)
  [arXiv:hep-ex/0212021];
  %%CITATION = PRLTA,90,021802;%%
  M.~H.~Ahn {\it et al.}  [K2K Collaboration],
  %``Indications of neutrino oscillation in a 250-km long-baseline experiment,''
  Phys.\ Rev.\ Lett.\  {\bf 90}, 041801 (2003)
  [arXiv:hep-ex/0212007];
  %%CITATION = PRLTA,90,041801;%%
  T.~Schwetz,
  %``Global fits to neutrino oscillation data,''
  Phys.\ Scripta {\bf T127}, 1 (2006)
  [arXiv:hep-ph/0606060];
  %%CITATION = PHSTB,T127,1;%%
  M.~Maltoni, T.~Schwetz, M.~A.~Tortola and J.~W.~F.~Valle,
  %``Status of global fits to neutrino oscillations,''
  New J.\ Phys.\  {\bf 6}, 122 (2004)
  [arXiv:hep-ph/0405172];
  %%CITATION = NJOPF,6,122;%%
  G.~L.~Fogli {\it et al.},
  %``Observables sensitive to absolute neutrino masses: A reappraisal after
  %WMAP-3y and first MINOS results,''
  Phys.\ Rev.\  D {\bf 75}, 053001 (2007)
  [arXiv:hep-ph/0608060].
  %%CITATION = PHRVA,D75,053001;%%
\bibitem{Ross:2000fn}
  G.~G.~Ross,
  ``Models of fermion masses,''
  %\href{http://www.slac.stanford.edu/spires/find/hep/www?irn=4868536}{SPIRES entry}
  {\it Prepared for Theoretical Advanced Study Institute in Elementary
  Particle Physics (TASI 2000): Flavor Physics for the Millennium,
  Boulder, Colorado, 4-30 Jun 2000}
\bibitem{mareview}
   E.~Ma,
  %``Non-Abelian Discrete Flavor Symmetries,''
  arXiv:0705.0327 [hep-ph].
  %%CITATION = ARXIV:0705.0327;%%
\bibitem{altarelli}
   G.~Altarelli,
  %``Models of neutrino masses and mixings: A progress report,''
  arXiv:0705.0860 [hep-ph].
  %%CITATION = ARXIV:0705.0860;%%
\bibitem{deMedeirosVarzielas:2006fc}
  I.~de Medeiros Varzielas, S.~F.~King and G.~G.~Ross,
  %``Neutrino tri-bi-maximal mixing from a non-Abelian discrete family
  %symmetry,''
  Phys.\ Lett.\  B {\bf 648}, 201 (2007)
  [arXiv:hep-ph/0607045].
  %%CITATION = PHLTA,B648,201;%%
\bibitem{A4papers}
   E.~Ma and G.~Rajasekaran,
  %``Softly broken A(4) symmetry for nearly degenerate neutrino masses,''
  Phys.\ Rev.\  D {\bf 64}, 113012 (2001)
  [arXiv:hep-ph/0106291];
  %%CITATION = PHRVA,D64,113012;%%
  K.~S.~Babu, E.~Ma and J.~W.~F.~Valle,
  %``Underlying A(4) symmetry for the neutrino mass matrix and the quark  mixing
  %matrix,''
  Phys.\ Lett.\  B {\bf 552}, 207 (2003)
  [arXiv:hep-ph/0206292];
  %%CITATION = PHLTA,B552,207;%%
  M.~Hirsch, A.~S.~Joshipura, S.~Kaneko and J.~W.~F.~Valle,
  %``Predictive flavour symmetries of the neutrino mass matrix,''
  arXiv:hep-ph/0703046.
  %%CITATION = HEP-PH/0703046;%%
\bibitem{frampton1}
  P.~H.~Frampton and T.~W.~Kephart,
  %``Simple nonAbelian finite flavor groups and fermion masses,''
  Int.\ J.\ Mod.\ Phys.\  A {\bf 10}, 4689 (1995)
  [arXiv:hep-ph/9409330].
  %%CITATION = IMPAE,A10,4689;%%
\bibitem{maximal}
   A.~Aranda, C.~D.~Carone and R.~F.~Lebed,
  %``Maximal neutrino mixing from a minimal flavor symmetry,''
  Phys.\ Rev.\  D {\bf 62}, 016009 (2000)
  [arXiv:hep-ph/0002044];
  %%CITATION = PHRVA,D62,016009;%%
  A.~Aranda, C.~D.~Carone and R.~F.~Lebed,
  %``U(2) flavor physics without U(2) symmetry,''
  Phys.\ Lett.\  B {\bf 474}, 170 (2000)
  [arXiv:hep-ph/9910392].
  %%CITATION = PHLTA,B474,170;%%
\bibitem{barbieri}
  R.~Barbieri, L.~J.~Hall, S.~Raby and A.~Romanino,
  Nucl.\ Phys.\ B{\bf 493}, 3 (1997)
  [arXiv:hep-ph/9610449].
\bibitem{feruglio}
   F.~Feruglio, C.~Hagedorn, Y.~Lin and L.~Merlo,
  %``Tri-bimaximal neutrino mixing and quark masses from a discrete flavour
  %symmetry,''
  Nucl.\ Phys.\  B {\bf 775}, 120 (2007)
  [arXiv:hep-ph/0702194].
  %%CITATION = NUPHA,B775,120;%%
\bibitem{muchunchen}
   M.~C.~Chen and K.~T.~Mahanthappa,
  %``CKM and Tri-bimaximal MNS Matrices in a SU(5) x (d)T Model,''
  arXiv:0705.0714 [hep-ph].
  %%CITATION = ARXIV:0705.0714;%%
\bibitem{frampton2}
  P.~H.~Frampton and T.~W.~Kephart,
  %``Flavor Symmetry for Quarks and Leptons,''
  arXiv:0706.1186 [hep-ph].
  %%CITATION = ARXIV:0706.1186;%%
\bibitem{muchun2}
   M.~C.~Chen
   arXiv:0706.2168v1 [hep-ph].
\end{thebibliography}
\end{document}